\documentclass{article}
\usepackage[utf8]{inputenc}
\usepackage{authblk}
\usepackage{setspace}
\usepackage[margin=1.25in]{geometry}
\usepackage{graphicx}
\graphicspath{ {./figures/} }
\usepackage{subcaption}
\usepackage{amsmath}
\usepackage{lineno}

\usepackage[style=nejm, 
citestyle=numeric-comp,
sorting=none]{biblatex}
\title{Toward the Stars: Technological, Ethical, and Sociopolitical Dimensions of Interstellar Exploration
}

\author[1,2*]{Florian Neukart}

\affil[1]{Leiden Institute of Advanced Computer Science, Leiden University, Snellius Gebouw, Niels Bohrweg 1, 2333 CA Leiden, Netherlands}
\affil[2]{Terra Quantum AG, Kornhausstrasse 25, 9000 St. Gallen, Switzerland}

\date{}

\begin{document}

\maketitle

\begin{abstract}
The endeavor of interstellar exploration is a convergence of technical innovation and profound ethical inquiry, challenging humanity to extend its reach beyond the confines of our solar system while contemplating the moral implications of such a leap. This paper explores the multifaceted aspects of interstellar travel, exploring advancements in propulsion systems, habitat construction, and life support alongside the ethical, sociopolitical, and philosophical questions that arise as we consider colonizing extraterrestrial worlds. We underscore the imperative for an integrative framework harmonizing scientific achievements with a deep ethical commitment to responsible exploration, environmental stewardship, and respect for potential extraterrestrial life. Our analysis highlights the dual nature of interstellar exploration as both a technical endeavor and a philosophical journey, advocating for a future in which humanity's expansion into the cosmos is guided by foresight, equity, and the collective well-being of all sentient beings. This synthesis of science and ethics offers a blueprint for navigating the unknowns of space with wisdom and integrity, ensuring that our interstellar aspirations reflect the best of human values.
\end{abstract}

\section{Introduction}
The advent of space exploration has ushered humanity into a new era of discovery and challenge. As we stand on the brink of becoming an interstellar species, this leap forward's technical, ethical, and philosophical dimensions demands rigorous examination. This paper is motivated by the growing feasibility of interstellar travel, driven by advancements in propulsion technology, life support systems, and sustainable habitats. Yet, as we chart courses to distant worlds, the need for a holistic approach to space colonization becomes apparent—one that encompasses not only the engineering and biological sciences but also addresses the profound ethical and sociopolitical questions that arise.

Interstellar exploration offers unparalleled opportunities for growth, discovery, and the expansion of human knowledge. However, it also presents unique challenges, including the ethical implications of terraforming, the governance of extraterrestrial colonies, and the preservation of potential extraterrestrial life forms. These considerations require a multidisciplinary approach, integrating insights from physics, engineering, biology, ethics, and social sciences to forge a technologically viable and morally sound path.

The structure of this paper reflects this integrative approach. Following this introduction, we review the current technology state and theoretical frameworks underpinning interstellar travel in the \textit{Technical Foundations} section. We then proceed to \textit{Ethical Considerations}, examining the moral dilemmas posed by extending human presence beyond Earth. The \textit{Sociopolitical Implications} section explores the governance and societal impacts of space colonization. In contrast, \textit{Philosophical and Evolutionary Perspectives} elaborates on the existential questions and potential evolutionary trajectories for humans in space. Finally, we discuss the challenges and opportunities that lie ahead, asserting the importance of ethical stewardship and innovative thinking in navigating humanity's future among the stars.

\section{Background}
The endeavor to extend human presence beyond Earth has evolved from the early days of space exploration to ambitious plans for Mars colonization. The technological advancements in propulsion, habitat construction, and life support systems have transformed interstellar travel from science fiction to a foreseeable future. Concurrently, discussing such ventures' ethical, sociopolitical, and environmental implications has gained momentum, underscoring the need for a multidisciplinary approach to space colonization.

\subsection{Technological Advancements}
The quest for interstellar exploration has catalyzed significant advancements in space technology, with propulsion systems, life support, and habitat construction at the forefront of research and development. These innovations are critical for overcoming space's immense distances and harsh conditions, enabling sustainable human presence beyond Earth.

\subsubsection{Advancements in Propulsion Technologies}
\paragraph{Chemical Rockets:} Traditional chemical rockets, while effective for Earth-to-orbit missions, are severely limited by their thrust-to-weight ratio and the tyranny of the rocket equation. The need for vast propellants to achieve even a fraction of light speed renders them impractical for interstellar travel \cite{Nerem2001A}.

\paragraph{Magnetic Fusion Plasma Drive (MFPD):} A revolutionary propulsion concept, the MFPD leverages the immense energy potential of nuclear fusion to achieve unprecedented thrust and efficiency. This approach involves the magnetic confinement of fusion plasma, offering a scalable solution for powering large spacecraft over interstellar distances. By utilizing deuterium and tritium as fuel, the MFPD promises a significant advancement in propulsion technology, potentially reducing travel times to nearby star systems from millennia to mere decades. The successful implementation of MFPD technology would represent a monumental leap forward, making interstellar travel feasible \cite{neukart2023magnetic, Winterberg2000Fusion}.

\paragraph{Ion Thrusters:} Ion thrusters represent a significant leap forward, offering higher specific impulse (Isp) than chemical rockets by ionizing and accelerating propellant, typically xenon, using electric power. While current models produce low thrust, their efficiency allows for gradual acceleration, making them suitable for long-duration, deep-space missions \cite{Choueiri2009Ion, Sanchez2018Recent, neukart2023magnetic}.

\paragraph{Nuclear Thermal and Electric Propulsion:} Nuclear propulsion offers two main variants—thermal and electric. Using nuclear reactions, nuclear thermal propulsion (NTP) heats a propellant, like hydrogen, achieving higher Isp than conventional chemical rockets. Nuclear electric propulsion (NEP) converts nuclear energy into electrical power, which drives an electric propulsion system, such as an ion thruster, providing even greater efficiency over long distances \cite{Emrich2012Overview, Brophy2016Nuclear, neukart2023magnetic}.

\paragraph{Breakthrough Starshot Initiative:} Perhaps the most ambitious concept, the Breakthrough Starshot initiative, proposes using powerful ground-based lasers to propel lightweight spacecraft with light sails to a significant fraction of the speed of light. This method could allow probes to reach the nearest star systems, such as Alpha Centauri, in just a few decades \cite{Lubin2016Towards}.

\paragraph{Integration of Multidisciplinary Approaches:} Building on the principles highlighted in the MFPD \cite{neukart2023magnetic} research and sustainable Mars colonization strategies \cite{neukart2024towards}, it's evident that the future of interstellar propulsion technologies relies not only on advancements in engineering but also on a multidisciplinary approach encompassing material science, environmental ethics, and economic feasibility. This integrated perspective ensures the development of propulsion systems that are not only technologically advanced but also sustainable and ethically sound for long-term space exploration and colonization efforts.

\paragraph{Innovations in Resource Utilization:} The exploration into in-situ resource utilization (ISRU) for Mars colonization \cite{neukart2024towards} underscores the importance of leveraging local resources for propulsion technologies. This approach could significantly reduce the logistical and economic challenges associated with interstellar travel, highlighting the potential for developing propulsion systems that utilize resources harvested directly from space environments, thus enhancing the sustainability and autonomy of interstellar missions.

\paragraph{Enhanced Focus on Environmental Sustainability:} The research into Martian concrete and sustainable habitat construction \cite{neukart2024towards} emphasizes the critical role of environmental sustainability in space exploration. Similarly, propulsion technologies should prioritize minimal environmental impact in space and on celestial bodies. This includes reducing the reliance on non-renewable resources and exploring propulsion methods that minimize space debris and other forms of cosmic pollution, ensuring that humanity's leap into interstellar space is responsible and sustainable.

\subsubsection{Closed-Loop Life Support Systems}
The sustainability of human life in space environments, void of Earth's natural life support, necessitates the development of closed-loop life support systems (CLLSS). These systems are engineered to recycle water, air, and waste with minimal external input, mimicking Earth's ecological cycles to sustain life. Advances in bioregenerative life support systems (BLSS) integrate complex biological processes, leveraging the capabilities of algae and higher plants for oxygen production and carbon dioxide removal through photosynthesis. Innovations from sustainable Mars colonization efforts highlight the potential for genetically engineered plants and microorganisms, enhancing the efficiency of these processes \cite{neukart2024towards, Hendrickx2005Overview, Koenig2014Advanced}. Advanced water recycling techniques, inspired by Martian water extraction methods, further optimize CLLSS, ensuring the viability of long-duration missions and extraterrestrial colonies \cite{neukart2024towards, Hendrickx2005Overview}.

\paragraph{Water Recycling:} Water recycling in CLLSS involves a series of filtration and purification processes, including reverse osmosis, bioreactors for organic matter breakdown, and electrolysis for disinfection. These processes ensure the continuous availability of clean water from waste fluids, including urine and humidity from the air \cite{Hendrickx2005Overview}.

\paragraph{Atmosphere Regeneration:} Atmosphere regeneration is achieved through a combination of physical-chemical and biological methods. Carbon dioxide scrubbers use materials like lithium hydroxide to remove CO$_2$ from the air chemically. At the same time, plants and photosynthetic algae in BLSS convert CO$_2$ back into oxygen, enhancing air quality and supporting the carbon cycle \cite{neukart2024towards, Hendrickx2005Overview}.

\paragraph{Food Production and Waste Recycling:} BLSS also explores integrating hydroponic and aquaponic systems for food production, utilizing waste materials as nutrients for plant growth. These systems provide fresh produce to supplement astronauts' diets and contribute to crew members' psychological well-being by introducing greenery and a semblance of Earth's environment \cite{neukart2024towards, Hendrickx2005Overview}.

\paragraph{Psychological Benefits:} The inclusion of bioregenerative components in life support systems offers psychological benefits by creating a more natural and comforting living space. Studies suggest that interacting with plant life can reduce stress and improve mental health, an essential aspect of long-duration space missions \cite{neukart2024towards, Hendrickx2005Overview}.

The development and optimization of CLLSS and BLSS are critical for future long-duration space missions, interplanetary travel, and permanent off-Earth settlements. These systems represent a convergence of engineering, biology, and environmental science, requiring multidisciplinary research to address the challenges of closed-environment living and resource sustainability in space.

\subsubsection{3D Printing Technologies for Habitat Construction}
Constructing habitable structures on other planets presents unique challenges, primarily the logistical difficulty of transporting building materials from Earth and the harsh environmental conditions of extraterrestrial surfaces. 3D printing technology, especially when combined with ISRU, offers a promising solution to these challenges. Utilizing local materials, such as lunar regolith or Martian soil, for printing habitat components significantly reduces reliance on Earth-based resources, lowering costs and enhancing the feasibility of building durable, radiation-shielded living spaces. Insights from the development of Martian concrete suggest significant advancements in material science, enabling the creation of regolith-based concrete with optimized mechanical properties for extraterrestrial construction \cite{neukart2024towards, Khoshnevis2004Automated, Lim2011Lunar}. The adaptation of 3D printing technology for these environments involves overcoming engineering challenges, with innovative solutions including robotic printers and autonomous construction systems, promising to revolutionize the construction of extraterrestrial habitats \cite{Khoshnevis2004Automated, Lim2011Lunar}.

\paragraph{Material Innovation and ISRU:} The development of regolith-based concrete represents a significant advancement in material science for extraterrestrial construction. This innovation involves processing and combining lunar or Martian soil with binders to produce a concrete-like material suitable for 3D printing. The mechanical properties of regolith-based concrete, including compressive strength and radiation shielding capabilities, are optimized for the demanding conditions of space habitats \cite{neukart2024towards, Khoshnevis2004Automated, Lim2011Lunar}.

\paragraph{Engineering Challenges and Solutions:} The adaptation of 3D printing technology for extraterrestrial environments requires overcoming several engineering hurdles, such as the fine-tuning of print parameters to accommodate the unique properties of regolith-based materials and the development of printers capable of operating in reduced gravity and extreme temperatures. Innovative solutions, including robotic printers and autonomous construction systems, are under development to address these challenges, aiming to enable the efficient assembly of structures on planetary surfaces without direct human intervention \cite{Khoshnevis2004Automated, Lim2011Lunar}.

\paragraph{Demonstrations and Future Prospects:} Recent demonstrations of 3D-printed structures using simulated lunar and Martian materials have validated the feasibility of this approach. These prototypes have showcased the potential for creating various structures, from simple habitats to complex infrastructural elements, capable of withstanding the harsh conditions of space. The continuous refinement of 3D printing technologies and materials promises to revolutionize the construction of extraterrestrial habitats, making a sustainable human presence on other planets increasingly attainable \cite{Khoshnevis2004Automated, Lim2011Lunar}.

Integrating 3D printing and ISRU represents a pivotal advancement in space exploration technologies, offering a scalable and efficient method for constructing habitable environments beyond Earth. This approach addresses the logistical and environmental challenges of extraterrestrial construction and paves the way for future missions to establish permanent human settlements in space.

\subsection{Ethical and Sociopolitical Frameworks}
The venture into outer space presents multifaceted ethical dilemmas and sociopolitical challenges, magnifying the need for comprehensive legal and ethical frameworks. The Outer Space Treaty of 1967 laid foundational principles for space exploration, emphasizing the peaceful use of outer space and declaring space as the "province of all mankind." However, as we edge closer to establishing a permanent human presence beyond Earth, questions regarding the rights of potential extraterrestrial life forms, the appropriation and use of space resources, and the governance of extraterrestrial colonies demand more detailed resolutions \cite{OuterSpaceTreaty}.

\paragraph{Legal Frameworks and Governance Issues:} Recent discussions within the United Nations Committee on the Peaceful Uses of Outer Space (COPUOS) and other international forums have highlighted the inadequacies of current treaties in addressing the complexities of space colonization. Issues such as space resource mining, territory claims on celestial bodies, and the autonomy of space colonies necessitate the development of new legal instruments and governance models that ensure equitable and sustainable use of space resources, respect for potential extraterrestrial ecosystems, and fair governance structures for space communities \cite{Weeden2017Survey}.

\paragraph{Ethical Considerations in Interstellar Exploration:} The ethical considerations of interstellar exploration extend beyond legal frameworks, touching upon the moral obligations humans have towards each other, extraterrestrial life, and future generations. Debates surrounding planetary protection, the ethics of terraforming, and the potential impact on indigenous space ecosystems underscore the need for an ethical approach to space exploration that prioritizes preserving life and the integrity of celestial environments \cite{Cockell2015Astrobiology}.

\subsection{Philosophical Considerations}
Embarking on a journey to become an interstellar species compels humanity to confront profound philosophical questions concerning our identity, purpose, and destiny in the universe. The possibility of encountering extraterrestrial intelligence, the implications of long-term space habitation on human evolution, and the existential risks associated with space exploration force a reevaluation of our place in the cosmos.

\paragraph{Humanity's Place in the Universe:} The search for extraterrestrial life and the potential discovery of intelligent civilizations challenge human-centric views of the universe, prompting reflections on our responsibilities as cosmic citizens and stewards of life. Philosophical debates on the significance of expanding human presence into space, preserving diverse forms of life, and the ethical implications of altering extraterrestrial ecosystems highlight the complex interplay between human aspirations and cosmic responsibilities \cite{Vakoch2013Astrobiology}.

\paragraph{Existential Risk and Interstellar Legacy:} The long-term survival of humanity and the preservation of Earth's biosphere emerge as central themes in discussions on existential risk. The prospect of interstellar colonization offers a pathway to mitigate risks such as asteroid impacts and supernova events. Yet, it also raises questions about the legacy humanity wishes to leave in the universe, including the ethical implications of spreading life beyond Earth and the potential consequences of altering the course of cosmic evolution \cite{Bostrom2003Astronomical}.

This expanded discussion on ethical, sociopolitical, and philosophical frameworks sets the stage for a deeper exploration of interstellar ethics, integrating the technical challenges of space exploration with broader considerations of morality, governance, and human destiny. The subsequent sections build on this foundation, proposing a comprehensive approach to navigating the multifaceted dimensions of humanity's future among the stars.

\section{Technical Foundations}
The leap to interstellar space requires breakthroughs in propulsion, sustainable life support, and habitat construction. This section discusses the current technologies and theoretical proposals underpinning our aspirations for interstellar exploration, highlighting the scientific and engineering challenges that must be addressed.

\subsection{Propulsion Technologies}
Interstellar travel presents formidable challenges, necessitating propulsion technologies far beyond the capabilities of current chemical rockets. This section explores advanced propulsion systems, including nuclear thermal propulsion, ion drives, theoretical warp drives, and the Magnetic Fusion Plasma Drive (MFPD), focusing on their operational principles, energy requirements, and the physics governing their potential and limitations \cite{Finney2017Interstellar, Millis2014Frontiers, neukart2023magnetic}.

\subsubsection{Magnetic Fusion Plasma Drive (MFPD)}
The MFPD \cite{neukart2023magnetic} is a propulsion concept designed to harness the immense power of nuclear fusion for interstellar travel. By utilizing magnetically confined plasma from fusion reactions, the MFPD aims to overcome the limitations of current propulsion methods, offering a scalable solution with unparalleled efficiency and thrust capabilities.

\paragraph{Operational Principles:}
At the heart of the MFPD lies a fusion reactor where isotopes of hydrogen, deuterium, and tritium are fused at high temperatures, releasing vast amounts of energy. This process generates high-energy plasma for propulsion and produces neutrons which can be harnessed for additional power generation, making the MFPD a dual-purpose system.

\paragraph{Advantages and Scalability:}
The MFPD boasts several advantages over traditional propulsion systems, including significantly higher specific impulse and the ability to provide continuous thrust over long durations. Its scalability suits various mission profiles, from cargo transport to crewed interstellar voyages. Using abundant isotopes like deuterium as fuel and the potential for self-sustaining power generation through fusion reactions underscore the MFPD's role in future space exploration.

\paragraph{Challenges and Future Outlook:}
Despite its promising potential, the MFPD faces technical challenges, particularly in achieving controlled fusion in space and managing the intense heat and pressures within the reactor. Advances in materials science and magnetic confinement techniques are critical to realizing the full potential of MFPD technology. Ongoing research and development efforts are focused on addressing these challenges to make interstellar travel a reality.

\paragraph{Implications for Interstellar Travel:}
The development of the MFPD represents a significant milestone in our quest to become an interstellar species. Its successful implementation could dramatically reduce travel times to nearby star systems, opening new horizons for human exploration and the potential colonization of exoplanets. The MFPD promises a leap forward in propulsion technology and offers a glimpse into a future where interstellar travel is within our reach \cite{neukart2023magnetic}.

\subsubsection{Nuclear Thermal Propulsion (NTP)}
Nuclear Thermal Propulsion represents a distinct branch of propulsion technology, leveraging nuclear fission to heat a propellant, such as hydrogen, which then expands and is expelled to produce thrust. This method significantly improves specific impulse (Isp) over chemical rockets, achieving values between 850 to 1000 seconds, nearly double that of the most efficient chemical systems. NTP's higher efficiency and thrust capabilities make it particularly valuable for missions requiring significant delta-v, such as crewed Mars missions or other deep space explorations \cite{Finney2017Interstellar, neukart2023magnetic}.

\paragraph{Mechanism and Operation:}
The core principle of NTP involves a nuclear reactor that initiates a controlled fission reaction, releasing heat. This heat is transferred to the propellant, causing it to expand and be expelled through a nozzle, creating thrust akin to traditional chemical rockets but at a higher efficiency due to the greater exhaust velocities achievable with nuclear fission compared to chemical reactions.

\paragraph{Advantages and Potential:}
NTP systems are characterized by their high thrust-to-weight ratio and a significant reduction in travel time for deep space missions. The use of hydrogen as a propellant not only provides high specific impulse but also reduces the spacecraft's overall launch mass compared to chemical propulsion systems \cite{neukart2023magnetic}.

\paragraph{Limitations and Challenges:}
Key challenges include managing the radiation produced by the nuclear reactor and requiring robust shielding to protect both the spacecraft's systems and crew. The technical complexity of handling fissile material and controlling nuclear reactions also presents considerable engineering hurdles. Environmental and safety concerns, along with political and regulatory issues, also play a significant role in deploying NTP technologies in space \cite{neukart2023magnetic}.

\subsubsection{Ion Drives}
\paragraph{Physics and Operational Principles:}
Ion drives, a class of electric propulsion systems, utilize electricity—often from solar panels or nuclear reactors—to ionize a propellant (typically xenon or argon), and then accelerate it to high velocities using electromagnetic fields. This process generates thrust, propelling the spacecraft forward. The efficiency of ion drives is characterized by their high specific impulse (Isp), which can exceed 3000 to 5000 seconds, significantly surpassing the capabilities of traditional chemical rockets \cite{Millis2014Frontiers, neukart2023magnetic}.

The operational principle of ion drives centers on the acceleration of ions created from the onboard propellant. These ions are expelled at extremely high velocities, which allows for the gradual increase of spacecraft velocity over time despite the low thrust produced. This makes ion drives particularly suitable for deep-space missions, where high delta-v maneuvers are required over extended periods \cite{neukart2023magnetic}.

\paragraph{Advantages over Traditional Propulsion Systems:}
Ion drives offer several advantages over chemical propulsion systems, including significantly higher efficiency and lower propellant requirements. This efficiency is measured in specific impulses, with ion drives achieving values many times greater than the most efficient chemical rockets. Such high efficiency translates to reduced propellant mass, extending mission durations and enabling complex interplanetary trajectories that would be impractical with conventional propulsion \cite{neukart2023magnetic}.
Moreover, compared to other advanced propulsion concepts like the MFPD, ion drives present a more mature technology, currently deployable for missions requiring sustained thrust over long durations. However, they lack the high thrust-to-weight ratio and the potential for rapid interplanetary travel offered by fusion-based systems like the MFPD, which promise revolutionary advances in space propulsion through the use of nuclear fusion to achieve unprecedented thrust levels and efficiency \cite{neukart2023magnetic}.

\paragraph{Limitations and Challenges:}
While ion drives excel in efficiency and endurance, they produce relatively low thrust levels, making them unsuitable for missions that require quick acceleration or heavy lifting capabilities from planetary surfaces. The reliance on electrical power also necessitates substantial power generation and thermal management systems, adding complexity and potentially increasing the spacecraft's mass. Additionally, the technology's dependence on rare or expensive propellants and the erosion of thruster components over time pose operational and logistical challenges that must be addressed to optimize performance and longevity \cite{neukart2023magnetic}.

Ion drives are a significant advancement in space propulsion technology, offering high efficiency and reduced propellant consumption for long-duration missions. However, the exploration of more powerful and efficient propulsion methods, such as the MFPD, continues to be crucial for the future of interplanetary and potentially interstellar travel \cite{Millis2014Frontiers, neukart2023magnetic}.

\subsubsection{Theoretical Possibilities of Warp Drives}
Warp drives represent a theoretical framework for faster-than-light (FTL) travel, fundamentally challenging our current understanding of physics and propulsion. Rooted in solutions to Einstein's general relativity equations, warp drives, notably the concept proposed by Alcubierre, suggest the possibility of bending or warping spacetime around a spacecraft, creating a bubble that allows for superluminal travel without violating the speed of light within the bubble's local frame.

\paragraph{Energy Requirements and Theoretical Models:}
The Alcubierre drive model requires a form of exotic matter with negative energy density to distort spacetime. This matter, not observed in nature, would allow the expansion of spacetime behind the spacecraft and its contraction in front, propelling it forward at effective speeds surpassing light. Initial estimates suggested prohibitive amounts of energy, equivalent to the mass-energy of planets or even stars, posing significant challenges for practical implementation \cite{Alcubierre1994, Lobo2007}.

\paragraph{Recent Advances and Reduction in Energy Requirements:}
Recent theoretical advancements have proposed modifications to the original Alcubierre metric, significantly reducing the energy requirements for creating and maintaining a warp bubble. These include altering the geometry of the bubble and utilizing quantum effects to stabilize the exotic matter needed. Such developments have sparked renewed interest in the feasibility of warp drives, although practical energy requirements remain exceedingly high for current technological capabilities \cite{White2013, Lentz2020}.

\paragraph{Challenges and Implications for Interstellar Travel:}
Beyond the energy considerations, the realization of warp drives faces numerous theoretical and engineering challenges. These include creating and stabilizing exotic matter, protecting the spacecraft and its occupants from harmful radiation and spacetime distortions, and the potential impact on destination points due to the release of high-energy particles upon bubble collapse. Moreover, the implications of causality and the possibility of time travel introduce further complexity into the theoretical framework of warp drives \cite{Everett2012, Lobo2007}.

Despite these challenges, the theoretical exploration of warp drives offers valuable insights into the limits of our understanding of spacetime, propulsion, and the potential for future breakthroughs in interstellar travel. As research progresses, the concept of warp drives remains a fascinating and speculative area of study, representing the outer bounds of what might be possible in the quest to traverse the vast distances of the cosmos \cite{Alcubierre1994, White2013, Lentz2020}.

This exploration of advanced propulsion technologies underscores the critical role of innovative propulsion methods, like the MFPD, in realizing the dream of interstellar exploration.

\subsection{Life Support and Habitat Construction}
The challenge of sustaining human life throughout interstellar voyages and within extraterrestrial environments necessitates the development of advanced life support systems and construction techniques. These systems must efficiently recycle air, water, and nutrients, while habitat construction must utilize available resources to ensure human survival in harsh extraterrestrial conditions.

\subsubsection{Closed-Loop Life Support Systems}
Engineering robust closed-loop life support systems involves integrating advanced technologies to recycle water, air, and waste. These systems employ biological and physicochemical processes to mimic Earth's natural life support cycles. For example, water recovery systems use a combination of filtration, reverse osmosis, and catalytic oxidation to purify wastewater for reuse. Air revitalization systems incorporate carbon dioxide scrubbers and electrolysis units to split water into oxygen and hydrogen, maintaining breathable air quality.

\paragraph{Mathematical Modeling of Life Support Systems:}
The efficiency and reliability of closed-loop life support systems are enhanced through mathematical modeling and simulation. Models help predict system behavior under various conditions, optimizing resource recycling and minimizing waste. For instance, the mass balance equation, \(m_{in} - m_{out} + m_{gen} - m_{cons} = m_{acc}\), where \(m\) represents mass flows, is critical for designing systems that balance the input, output, generation, and consumption of resources \cite{Jones2008Mathematical}.

\subsubsection{Habitat Construction Utilizing ISRU}
Utilizing in-situ resources for habitat construction on alien worlds requires innovative engineering approaches and materials science. ISRU-based construction techniques, such as 3D printing with regolith-derived materials, offer promising solutions for building durable structures that protect inhabitants from extreme temperatures, radiation, and micrometeorite impacts \cite{Rygalov2013Regolith, Lim2018Lunar}.

\paragraph{Engineering Challenges and Solutions:}
Developing ISRU technologies for habitat construction faces several engineering challenges, including material processing, fabrication in reduced gravity conditions, and integrating habitats with life support systems. Advances in robotics and autonomous construction methods are critical for overcoming these obstacles, enabling the remote building of habitats before human arrival \cite{Bosman2017Automation}.

\paragraph{Structural Integrity and Radiation Shielding:}
The design of extraterrestrial habitats must consider structural integrity and radiation shielding. Engineering analyses, such as finite element modeling (FEM), assess the strength and stability of structures under extraterrestrial conditions. Materials science is crucial in developing regolith-based composites with enhanced radiation shielding properties, which is critical for long-term human safety \cite{Paz2018Structural}.

The advancement of life support and habitat construction technologies is pivotal for interstellar exploration and colonization success. These endeavors require a multidisciplinary approach, combining engineering, physics, biology, and materials science to solve the complex challenges of sustaining human life in space.

\subsection{Challenges and Opportunities}
The journey towards realizing interstellar travel encompasses a spectrum of challenges, ranging from the vast energy demands to the intricacies of navigating and surviving in deep space environments. These challenges are not merely obstacles but catalysts for innovation, pushing the frontiers of physics, engineering, biology, and more toward unprecedented discoveries.

\subsubsection{Energy Requirements and Propulsion Challenges}
One of the paramount hurdles is the immense energy required for propelling spacecraft across interstellar distances. Theoretical propulsion systems, such as the Magnetic Fusion Plasma Drive or antimatter engines, promise significant advancements but require energy generation and containment breakthroughs. The physics of achieving and managing such high-energy outputs while ensuring spacecraft integrity and crew safety demands rigorous research and innovative engineering solutions \cite{Forward1995Roundtrip, Phipps2000Starship}.

\paragraph{Mathematical Formulation:}
The energy requirements for interstellar propulsion can be estimated through relativistic equations, considering the mass-energy equivalence principle, \(E=mc^2\), and the specific impulse needed to achieve near-light-speed travel. Detailed calculations must account for the spacecraft's mass, desired velocity, and propulsion system's efficiency, guiding the development of feasible interstellar travel technologies \cite{Matloff2010Deep}.

\subsubsection{Navigating Deep Space Environments}
Deep space presents a myriad of unknowns, from the effects of prolonged exposure to cosmic radiation on human health to the potential for encountering interstellar matter that could impact spacecraft integrity. Developing reliable navigation systems that can adjust for these unpredictable factors is crucial. This entails advancements in spacecraft shielding, medical research, and autonomous navigation algorithms capable of real-time decision-making in uncharted territories \cite{Kanas2008Space, Geller2016Review}.

\paragraph{Engineering and Biological Innovations:}
Addressing these challenges requires a multi-disciplinary approach, incorporating advanced materials science for radiation shielding, biotechnology for health maintenance, and artificial intelligence for navigation and system management. Each of these areas must be explored with an eye toward the unique demands of interstellar travel, balancing scientific ambition with practical feasibility \cite{Cortesao2019Space}.

Transitioning from theoretical models to practical interstellar travel technologies underscores a dynamic interplay between challenges and opportunities. As we push the boundaries of what is technologically possible, we also broaden our understanding of the universe and our place within it. This technical foundation propels us towards new horizons in space exploration and deepens our grasp of the ethical, sociopolitical, and philosophical implications of extending humanity's reach into the cosmos.

\section{Ethical Considerations}
The ethical implications of such a monumental step cannot be overstated, as humanity stands on the threshold of becoming an interstellar species. This section explores the ethical considerations surrounding interstellar exploration and colonization, emphasizing the need for a framework that balances human aspirations with the stewardship of the cosmos.

\subsection{The Ethics of Terraforming and Environmental Stewardship}
Terraforming, the deliberate alteration of extraterrestrial environments to support human life, presents profound ethical challenges. These challenges revolve around the potential existence of indigenous life forms, the preservation of untouched celestial landscapes, and the broader implications of imposing significant ecological changes on a planetary scale. We explore these concerns within the frameworks of environmental ethics, specifically through the lenses of bio-centrism and eco-centrism.

\subsubsection{Bio-centrism and Indigenous Life Forms}
Bio-centrism argues for the intrinsic value of all living entities, positing that the rights of potential indigenous extraterrestrial life should be a primary consideration in terraforming decisions. This perspective raises critical questions about our moral obligations to preserve and protect alien life forms, even at the microbial level, challenging the anthropocentric view that human interests supersede those of other species \cite{Cockell2000Ethical}.

\subsubsection{Eco-centrism and Celestial Ecosystems}
Eco-centrism extends ethical consideration to entire ecosystems, emphasizing the value of geological and microbial systems as integral components of a planet's natural heritage. From this viewpoint, terraforming efforts must weigh the impact on these systems, acknowledging the potential loss of pristine extraterrestrial environments and the unforeseeable consequences of ecological alterations on a planetary scale \cite{Baum2018Ecoethical}.

\subsubsection{Ethical Frameworks for Decision Making}
Applying bio-centric and eco-centric principles to terraforming initiatives necessitates a comprehensive ethical framework that guides decision-making processes. This framework should account for the preservation of extraterrestrial ecosystems, respect for potential indigenous life, and the prudent use of technology in altering planetary environments. Engaging in thoughtful, ethical discourse, incorporating a wide range of perspectives, is crucial for navigating the moral landscape of terraforming and ensuring responsible stewardship of the cosmos \cite{Singer2016Ethical}.

In examining the ethics of terraforming and environmental stewardship, it becomes evident that the endeavor to make alien worlds habitable for humans intersects deeply with broader ethical, philosophical, and environmental considerations. The principles of bio-centrism and eco-centrism serve as critical guides in this exploration, challenging us to consider the far-reaching implications of our actions on extraterrestrial environments and their potential inhabitants.

\subsection{Resource Utilization and Interstellar Equity}
The exploration and utilization of extraterrestrial resources raise complex ethical questions, central to which is the principle of interstellar equity. This principle challenges us to consider how resources beyond Earth can be used responsibly, ensuring that their benefits are distributed equitably among all of humanity rather than accruing to a privileged few.

\subsubsection{Equitable Distribution of Space Resources}
Equitable distribution requires mechanisms to ensure that the wealth generated from extraterrestrial resources contributes to global development rather than exacerbating existing inequalities. This entails international cooperation and regulatory frameworks prioritizing shared scientific advancement and socioeconomic benefits across global communities \cite{Launius2005Aerospace}.

\subsubsection{Avoiding Interstellar Colonialism}
The history of colonialism on Earth provides a cautionary tale for space exploration. Avoiding interstellar colonialism involves respecting the integrity of celestial bodies and potential indigenous life forms and fostering a culture of exploration that is inclusive, ethical, and respectful of all stakeholders, including future generations \cite{Miller2015Making}.

\subsubsection{Implications for Earth's Environment and Economies}
The extraction and utilization of space resources must also consider the environmental impact on Earth, ensuring that space activities do not contribute to the degradation of our home planet's ecosystems. Furthermore, the economic implications of accessing potentially unlimited space resources include the potential for disruptive impacts on global markets and industries, necessitating careful management to prevent economic destabilization \cite{Vakoch2019Ecoethical}.

\subsubsection{Guidelines for Sustainable Space Resource Use}
Developing guidelines for the sustainable use of space resources involves integrating principles of environmental stewardship, economic fairness, and social responsibility. These guidelines should support the advancement of space technologies and exploration while ensuring that the benefits and risks are managed with a commitment to equity, sustainability, and the well-being of all Earth's inhabitants \cite{Smith2005Environmental}.

In exploring the concept of interstellar equity, it becomes clear that the ethical use of extraterrestrial resources is not only a matter of technological and scientific innovation but also of moral imperative. The sustainable and fair utilization of space resources offers a unique opportunity to address some of humanity's most pressing challenges, from environmental sustainability to social inequality, paving the way for a more equitable and prosperous future for all.

\subsection{Ethical Frameworks for Interstellar Societies}
The establishment of human societies in extraterrestrial environments presents unprecedented ethical challenges. These challenges revolve around governance, justice, human rights, and the potential to replicate Earth's historical injustices or forge societies embodying our highest values. Developing ethical frameworks for these new societies ensures they flourish while adhering to universal principles of justice and equity \cite{Zubrin2019The}.

\subsubsection{Governance Beyond Earth}
Effective governance in interstellar societies requires frameworks that promote transparency, accountability, and inclusivity. Drawing from democratic ideals and incorporating lessons learned from Earth's governance mistakes, these frameworks must facilitate equitable participation in decision-making processes, ensuring that all community members have a voice \cite{Scordato2016Extraterrestrial}.

\subsubsection{Justice and Human Rights}
The administration of justice in extraterrestrial communities must be founded on principles that guarantee fairness, respect for human dignity, and the protection of fundamental rights. Ethical considerations include adapting legal systems to the unique challenges of space environments, ensuring they are equipped to deal with issues of resource allocation, conflict resolution, and the safeguarding of individual freedoms \cite{Kluitenberg2019The}.

\subsubsection{Creating Equitable Societies}
To avoid the replication of Earth's social and economic inequalities, interstellar societies must prioritize the equitable distribution of resources and opportunities. This involves establishing economic systems that foster sustainable development, social welfare, and the eradication of poverty, while also respecting the limitations and potential vulnerabilities of extraterrestrial environments \cite{Cowan2019Interstellar}.

\subsubsection{Frameworks for Ethical Exploration and Settlement}
A commitment to peaceful coexistence, environmental stewardship, and the respectful treatment of any encountered extraterrestrial life should guide the development of ethical frameworks for interstellar societies. These frameworks must be flexible enough to evolve, incorporating new knowledge and adapting to the dynamic challenges of living beyond Earth \cite{Benford2012Starship}.

This exploration of ethical considerations underscores the importance of integrating moral reflection into the planning and executing interstellar exploration and colonization efforts. As we extend our reach into the cosmos, our actions must be guided by a commitment to do so responsibly, ensuring that our legacy among the stars is one of wisdom and respect for all forms of life and environments we encounter.

\section{Sociopolitical Implications}
Extending human activity into interstellar space carries profound sociopolitical implications, from the governance of extraterrestrial colonies to the impact on cultural and societal norms. This section explores these implications, emphasizing the need for forward-thinking policies that promote equity, democracy, and peaceful coexistence across the stars.

\subsection{Governance in Interstellar Human Societies}
The transition to establishing human colonies on other planets introduces profound challenges for governance. These challenges stem from the vast distances and isolation of these colonies and the intrinsic need for their significant autonomy. This scenario raises pivotal questions regarding the transplantability and adaptability of Earth-based governance models to the context of interstellar colonization \cite{Arnold2017Extraterrestrial}.

\subsubsection{Adapting Democratic Governance}
The principles of democratic governance, characterized by participation, representation, and accountability, must be re-envisioned to suit the spatial and social fabric of interstellar societies. Innovations in digital democracy and decentralized decision-making processes could play vital roles in ensuring that governance structures remain inclusive and responsive to the needs of space colony inhabitants \cite{Scharmen2019Space}.

\subsubsection{Ensuring the Rule of Law}
The rule of law, a cornerstone of equitable governance, faces unique implementation challenges in space colonies. Legal frameworks must be developed to address the specific conditions of space living, encompassing issues from resource allocation to interpersonal conflict while safeguarding individual rights and freedoms. Developing such frameworks requires a balance between universal human rights principles and the specificities of life in space environments \cite{Wetmore2018Mapping}.

\subsubsection{Human Rights in the Cosmos}
Protecting and promoting human rights in interstellar human societies necessitate a foundational commitment to the dignity and worth of every individual. This commitment must be enshrined in the constitutive documents of space colonies and reflected in their daily governance practices. Ensuring access to essential resources, freedom of expression, and protection from discrimination are imperative in fostering just and equitable spacefaring societies \cite{Gorman2016The}.

\subsubsection{Drawing on Historical Precedents and Political Theory}
The design of governance systems for interstellar colonization can benefit from carefully examining historical precedents of colonialism, autonomy, and federation on Earth alongside contemporary political theories advocating for pluralism, subsidiarity, and cosmopolitanism. These insights can guide the creation of governance models that prevent the replication of Earth's historical injustices and aspire to embody the highest aspirations of human society \cite{deLeon2018Humanity}.

Establishing governance structures in interstellar human societies presents an opportunity to rethink the principles and practices of governance. By drawing on historical experiences and contemporary political thought, humanity has the chance to forge governance systems that are robust, equitable, and adaptable to the extraordinary circumstances of living beyond Earth.

\subsection{Cultural and Societal Changes}
By its challenges and novelties, interstellar colonization is poised to alter human culture and societal constructs fundamentally. The isolation of extraterrestrial colonies and the necessity to adapt to unfamiliar environments and achieve self-sufficiency will catalyze the development of distinct cultural identities and novel social structures.

\subsubsection{Emergence of New Cultural Identities}
The unique conditions of life on distant worlds are expected to foster the emergence of new cultural identities, shaped by the specific challenges and experiences of space colonization. These identities will reflect the diverse backgrounds of space settlers and the innovative practices developed to thrive in extraterrestrial environments \cite{Vakoch2019Societal}.

\paragraph{Potential for Cultural Diversification:}
Cultural diversification in space colonies presents opportunities for enriching human culture and challenges maintaining a cohesive societal fabric. The synthesis of Earth-based cultures with the novel experiences of space living may give rise to vibrant, diverse communities that embody the resilience and adaptability of humanity \cite{Dick2010Sociology}.

\subsubsection{Risks of Fragmentation and Conflict}
While cultural diversification is a source of strength, it poses risks of societal fragmentation and conflict. The stark differences between life on Earth and in space colonies could exacerbate misunderstandings and tensions, necessitating deliberate efforts to foster intercultural dialogue and mutual respect \cite{Kopf2015Space}.

\paragraph{Strategies for Fostering Cultural Cohesion:}
Promoting cultural cohesion and mutual understanding among Earth-based and space-based communities requires proactive strategies. These might include exchange programs, shared cultural and educational initiatives, and the development of communication platforms designed to facilitate the exchange of ideas and experiences across vast distances \cite{Fiedler2016Intercultural}.

\subsubsection{Ensuring Social Integration and Equity}
Efforts to build cohesive and equitable interstellar societies must prioritize including diverse voices in governance and decision-making processes. Policies and practices that ensure equitable access to resources, opportunities, and representation will be critical in mitigating social stratification risks and ensuring all community members' harmonious integration \cite{Davis2019Social}.

The profound changes anticipated in the wake of interstellar colonization call for a nuanced understanding of the dynamics of cultural evolution and societal adaptation. By embracing diversity, equity, and inclusivity, humanity can navigate the challenges of establishing interstellar communities that reflect the best of human values, fostering a shared sense of identity and purpose across the stars.

\subsection{Ethical and Equitable Resource Distribution}
The equitable distribution of resources between Earth and its extraterrestrial colonies represents a paramount ethical challenge in the era of interstellar expansion. This challenge encompasses ensuring fair access to technology, fostering economic development that benefits all of humanity, and adhering to principles of environmental sustainability across celestial bodies.

\subsubsection{Access to Technology}
Equitable access to space technology is crucial for preventing a technological divide between Earth and space colonies. Policies must be implemented to facilitate the transfer of knowledge and technology, ensuring that advancements in space exploration and resource utilization benefit all human societies, not just a privileged few \cite{Farmer2017Equity}.

\subsubsection{Economic Development}
The economic implications of resource distribution from extraterrestrial ventures necessitate a framework supporting the sustainable development of Earth- and space-based communities. This involves the creation of economic policies that prioritize equitable growth, job creation, and avoiding exploitative practices that could deepen existing economic disparities \cite{Pawlowski2019Economic}.

\subsubsection{Environmental Sustainability}
Sustainable resource distribution must also consider the environmental impact on Earth and other celestial bodies. The principles of environmental stewardship should guide the extraction and utilization of space resources, ensuring that interstellar expansion does not compromise the ecological integrity of planets, moons, or asteroids \cite{Barensky2018Sustainability}.

\subsubsection{Policies for Fair Sharing}
Developing policies that ensure the fair sharing of the benefits and burdens of interstellar expansion is essential. These policies should be based on international cooperation and guided by ethical frameworks emphasizing justice, equity, and the common good. By engaging diverse stakeholders in policy-making, humanity can strive toward a future where interstellar expansion contributes to a more equitable and sustainable future for all \cite{Lee2020Global}.

The ethical considerations surrounding the distribution of resources in the context of interstellar colonization underscore the need for thoughtful and inclusive approaches to policy-making. By prioritizing equitable access to technology, sustainable economic development, and environmental stewardship, humanity can navigate resource distribution challenges to promote the well-being of all Earth's inhabitants and ensure the responsible stewardship of the cosmos.

The sociopolitical implications of interstellar exploration and colonization are vast and complex. As humanity embarks on this journey, we must do so with a commitment to creating inclusive, equitable, and sustainable societies, both on Earth and among the stars. By addressing these challenges proactively, we can ensure that our future in space is marked by peace, prosperity, and mutual respect.

\section{Philosophical and Evolutionary Perspectives}
The prospect of interstellar exploration and colonization not only challenges our technological capabilities but also prompts deep philosophical reflections and raises questions about the future evolution of the human species. This section discusses the existential implications of humanity's spread into the cosmos and the potential evolutionary trajectories that could emerge from life in extraterrestrial environments.

\subsection{Existential Reflections on Interstellar Expansion}
Interstellar exploration presents humanity with the unparalleled opportunity to address some of the most profound questions about our existence. The potential encounter with extraterrestrial life forms and possibly intelligent civilizations challenges our anthropocentric views and compels us to reconsider our place in the vast cosmic order. This subsection elaborates on these existential questions, leveraging insights from philosophy, theology, and the social sciences to explore the deep impact of interstellar expansion on human self-understanding.

\subsubsection{Challenging Anthropocentrism}
The discovery of extraterrestrial life would significantly challenge the anthropocentric perspective that has predominantly shaped human understanding of the universe. Such an encounter would force us to confront the possibility that we are not the only, nor the most significant, form of intelligent life, thereby prompting a radical reevaluation of our place and purpose in the cosmos \cite{Vakoch2019Astrobiology}.

\subsubsection{Redefining Identity and Purpose}
Interstellar expansion invites us to redefine our identity and purpose beyond terrestrial confines. Philosophical and theological inquiries into the nature of existence, ethics, and our responsibilities in a potentially inhabited universe become paramount as we navigate the moral landscape of interacting with other life forms \cite{Baum2011Preliminary}.

\subsubsection{The Destiny of Humanity in the Universe}
The prospect of extending our presence to other worlds also raises questions about humanity's destiny in the universe. Are we destined to spread life throughout the cosmos, or should our expansion be tempered by a profound respect for the cosmic order and the potential sanctity of extraterrestrial ecosystems? These questions challenge us to think deeply about the legacy we wish to create as interstellar travelers \cite{Kopf2021Cosmic}.

\subsubsection{Insights from Philosophy, Theology, and Social Sciences}
Drawing on philosophical concepts of existentialism, theological reflections on creation and stewardship, and social scientific analyses of human behavior and societal development, this exploration seeks to provide a multidimensional understanding of the implications of interstellar exploration. By engaging with these diverse disciplines, we can comprehensively view our responsibilities and potential as members of a broader cosmic community \cite{Baum2011Preliminary}.

Interstellar exploration compels us to confront fundamental existential questions, challenging us to consider our identity, purpose, and destiny in a universe that may be teeming with life. By reflecting on these questions, humanity can navigate the vast unknowns of space with a sense of humility, curiosity, and responsibility, ensuring that a deep respect for all forms of existence guides our journey among the stars.

\subsection{Human Evolution in Extraterrestrial Environments}
Adapting to life beyond Earth presents unprecedented challenges and opportunities for human evolution. The unique conditions of extraterrestrial environments, such as microgravity, exposure to cosmic radiation, and interaction with novel ecosystems, can potentially drive significant biological and cultural evolution. This could lead to the emergence of distinct human subspecies, raising profound questions about our identity and future in the cosmos.

\subsubsection{Biological Adaptations to Microgravity and Radiation}
Microgravity profoundly affects the human body, from muscle atrophy and bone density loss to changes in cardiac function and fluid distribution. Prolonged exposure to cosmic radiation poses additional risks, including increased cancer rates and potential genetic mutations. The evolutionary response to these conditions could lead to significant anatomical and physiological changes over generations, potentially resulting in divergent human phenotypes adapted to space living \cite{Crucian2018Space}.

\subsubsection{Cultural Evolution and Diversification}
Beyond biological evolution, the cultural and societal adaptations to life in space will play a critical role in shaping human communities. Isolation, reliance on technology, and the need for self-sufficiency could foster unique cultural identities and social structures, influencing everything from language and art to governance and social norms \cite{Kanas2015Humans}.

\subsubsection{Genetic Engineering and Directed Evolution}
The advent of genetic engineering technologies offers the possibility of directed human evolution, allowing for targeted adaptations to extraterrestrial living conditions. Ethical considerations surrounding genetic modifications, including the implications for diversity and equality, are paramount. The potential for creating "designer" traits to enhance adaptability to space environments opens up a complex debate on the future direction of human evolution \cite{Pence2018Genetic}.

\subsubsection{Scientific and Ethical Considerations}
Exploring the future of human adaptability and diversity in space necessitates a multidisciplinary approach, incorporating insights from evolutionary biology, genetics, space medicine, and ethics. The prospect of distinct human subspecies raises ethical questions about identity, equality, and our responsibilities to future generations. Balancing the pursuit of knowledge and advancement with ethical stewardship will be crucial as we navigate the unknowns of human evolution in the cosmos \cite{Harris2018Human}.

Human evolution in extraterrestrial environments invites us to contemplate our place in the universe and the pathways of our future evolution. As we embark on this journey, the interplay of biological and cultural evolution, underpinned by ethical considerations, will define the essence of humanity in the cosmos.

\subsection{Implications for Human Society and Culture}
The enduring human presence in extraterrestrial environments poses profound implications for societal norms, cultural expressions, and philosophical perspectives. As humanity extends its reach into the cosmos, the experiences and challenges of life in space are poised to inspire the creation of new myths, narratives, and philosophies, reflecting a deepened understanding of our existence, morality, and the essence of community.

\subsubsection{Evolution of Societal Norms}
The unique conditions of space habitation—ranging from microgravity to the necessity for communal interdependence in isolated colonies—will inevitably reshape societal norms. These changes may manifest in altered social structures, new forms of governance tailored to space communities, and evolved social contracts prioritizing collective well-being and sustainability in the face of limited resources.

\subsubsection{Cultural Expressions and New Narratives}
Interstellar exploration and colonization will catalyze a renaissance in cultural expression as artists, writers, and thinkers draw upon the vast expanse of the universe for inspiration. Creating art, literature, and music that capture space's awe, beauty, and isolation will enrich human culture and provide critical reflections on our place in the universe and the shared human condition.

\subsubsection{Philosophical Reflections and Ethical Inquiry}
The philosophical implications of venturing beyond Earth challenge us to reconsider fundamental questions of existence, ethics, and our responsibilities to other life forms and the environments we inhabit. The potential discovery of extraterrestrial life, in particular, would necessitate a profound ethical and philosophical reevaluation of our place in the cosmic order, urging a shift towards a more inclusive, universal outlook on life.

\subsubsection{Fostering a Culture of Curiosity and Resilience}
As societies evolve in response to the realities of space habitation, fostering a culture of curiosity, resilience, and ethical reflection becomes paramount. Encouraging continuous learning, adaptability, and a deep commitment to ethical principles will be essential for navigating the challenges of interstellar expansion, ensuring that humanity's future among the stars is marked by wisdom, compassion, and a relentless pursuit of knowledge.

The long-term implications of human life in space on society and culture are vast and complex. By embracing the changes and challenges that come with interstellar exploration, humanity can develop a richer, more diverse cultural heritage and a more profound understanding of its place in the cosmos.

The philosophical and evolutionary perspectives presented here underscore the transformative potential of interstellar exploration. As we venture beyond our solar system, the insights gained from reflecting on our place in the universe and adapting to new environments will be invaluable in guiding the evolution of human society and culture in a cosmic context.

\section{Challenges and Opportunities}
The journey towards becoming an interstellar species is fraught with challenges that span the scientific, ethical, and sociopolitical spectra. However, within these challenges lie significant opportunities for growth, discovery, and the advancement of human civilization. This section outlines the key obstacles and advantages, proposing pathways forward that honor our responsibilities to our species and the cosmos.

\subsection{Scientific and Technological Challenges}
The odyssey into interstellar space introduces formidable scientific and technological challenges. These challenges span the traversal of immense distances, the survival within harsh extraterrestrial environments, and the development of sustainable life support systems. Key areas requiring significant advancements include propulsion technologies, habitat engineering, biotechnology, and materials science. These fields are critical for surmounting the obstacles inherent in interstellar exploration.

\subsubsection{Propulsion Technologies}
Advancing propulsion technologies is essential for reducing travel times to feasible durations. Current chemical rockets are inadequate for interstellar missions due to their limited efficiency and the vast energy requirements for such journeys. Innovations in nuclear propulsion, such as nuclear thermal and nuclear electric propulsion, offer higher specific impulse and could significantly enhance mission viability. Additionally, theoretical concepts like the warp drive and antimatter propulsion represent frontier science that could revolutionize our approach to interstellar travel.

\subsubsection{Habitat Engineering}
Creating habitable environments in space or on other planets necessitates breakthroughs in habitat engineering. This involves the development of structures that can shield inhabitants from cosmic radiation, provide a stable atmosphere, and ensure access to water and food. Utilizing in-situ construction and life support resources minimizes dependency on Earth-based supplies, making long-duration missions feasible.

\subsubsection{Biotechnology and Life Support Systems}
Sustainable life support systems are vital for long-term human survival in space. Advances in biotechnology are key to developing closed-loop systems that recycle air, water, and waste. Engineering plants and microorganisms to support life in space habitats not only addresses the physical needs of astronauts but also contributes to psychological well-being by simulating an Earth-like environment.

\subsubsection{Materials Science}
The harsh conditions of space demand materials with exceptional properties. Innovations in materials science are required to create spacecraft and habitats that can withstand extreme temperatures, radiation, and the vacuum of space. Smart materials that adapt to changing environments and self-healing structures could play pivotal roles in ensuring the safety and longevity of space infrastructure.

The pursuit of solutions to these scientific and technological challenges not only propels humanity closer to achieving interstellar travel but also catalyzes technological innovation with broad applications on Earth. From energy efficiency and sustainable living to medical breakthroughs, the advancements driven by space exploration hold the potential to address some of the most pressing challenges facing our planet.

\subsection{Ethical and Sociopolitical Challenges}

Interstellar exploration propels humanity into uncharted territories, physically and through a maze of complex ethical dilemmas and sociopolitical challenges. The encounter with potentially habitable worlds and unknown ecosystems poses unprecedented ethical questions: What rights do these worlds and their potential inhabitants have? How do we balance our curiosity and desire for exploration with the imperative not to harm other life forms or ecosystems? The governance of interstellar communities and the equitable distribution of space resources further highlight the need for inclusive, equitable, and sustainable governance models. These challenges are obstacles and opportunities to reflect upon and embody the best of human values. They call for a collective effort to establish norms and principles that ensure the dignity and rights of all beings are respected, setting a precedent for future generations. This endeavor requires a multidisciplinary approach, incorporating insights from ethics, law, sociology, and political science, to forge governance structures capable of addressing interstellar existence's unique demands.

\subsection{Opportunities for Human Evolution and Cultural Enrichment}

The venture into interstellar space promises to drive biological evolution through adaptation to new environments and significant cultural and philosophical advancements. This journey is an unparalleled opportunity to diversify human perspectives, enrich our culture, and deepen our philosophical understanding of our place in the universe. As humanity adapts to life in space and on other planets, we will likely encounter challenges that stimulate cultural innovation and philosophical inquiry, fostering a society that values resilience, adaptability, and a profound appreciation for the diversity of life. Embracing these changes with an ethical and open-minded approach can amplify the richness of human civilization, enhancing our collective ability to navigate the future with wisdom and foresight. The cultural and philosophical innovations arising from interstellar exploration could inspire new forms of art, literature, and social organization, reflecting the breadth of human creativity and the depth of our existential contemplation.

\subsection{Forging a Path Forward}

The path to becoming an interstellar species is fraught with challenges but also replete with inspiring opportunities. By confronting these challenges with creativity, ethical integrity, and a steadfast commitment to scientific rigor, humanity can chart a course through the complexities of interstellar existence. This journey, undertaken with responsibility and a visionary outlook, promises to expand the frontiers of human knowledge, culture, and capability. As we navigate this path, we must do so with a deep respect for the cosmos and a commitment to fostering a future that honors our shared heritage as inhabitants of the universe. The journey towards interstellar exploration is a testament to human ingenuity and perseverance and an invitation to redefine our understanding of progress, community, and our place within the cosmos. By embracing this journey with a sense of ethical responsibility and collective aspiration, we can pave the way for a future that is not only technologically advanced but also culturally rich and philosophically profound.

\section{Conclusion}
The vision of humanity's future among the stars is based on technological prowess, ethical reflection, and sociopolitical wisdom. As we stand on the precipice of interstellar exploration, this paper has traversed the challenges and opportunities that define this monumental venture. From the engineering feats required to breach the vast distances of space to the moral imperatives of engaging with unknown ecosystems and the governance of future space societies, we have outlined a path that marries innovation with integrity. 

The essence of our journey into the cosmos transcends the mere act of discovery—it is a profound exercise in self-reflection, prompting us to reconsider our place in the universe and our responsibilities towards each other and potential otherworldly life. As we embark on this journey, we must do so with a commitment to ethical stewardship, ensuring that our legacy among the stars is marked by respect, compassion, and the pursuit of knowledge for the betterment of all.

In conclusion, interstellar exploration beckons us with a promise of unparalleled discovery and the opportunity to forge a future that is not only technologically advanced but also ethically enlightened and socially just. By embracing a holistic approach that integrates science, ethics, and governance, we can navigate the stars with the confidence that the light of human wisdom and the enduring values of our shared humanity guide our path. The cosmos awaits, not as a realm to conquer but as a domain to respectfully explore, learn from, and coexist within the endless quest for understanding and connection.

\printbibliography

\end{document}